\documentclass[final, aps, reprint, nofootinbib, citeautoscript, superscriptaddress, balancelastpage]{revtex4-1}
\usepackage[T2A, T1]{fontenc}

\usepackage{amsmath}
\usepackage{amssymb}
\usepackage{bm}

\usepackage{textcomp}
\usepackage[colorlinks, urlcolor=blue, citecolor=blue, linkcolor=blue, pdfstartview=FitH]{hyperref}
\usepackage[all]{hypcap}
\usepackage[dvipsnames]{xcolor}
\usepackage{subfigure}
\usepackage[pdftex]{graphicx}

\def\affiSOLAB{Spin\ Optics\ Laboratory, Saint~Petersburg\ State\ University, 198504 St.~Peterbsurg, Russia}

\begin{document}
\author{G.~G.\ Kozlov}
\author{A.~A.\ Fomin}
\author{M.~Yu.\ Petrov}
\affiliation{\affiSOLAB}
\author{V.~S.\ Zapasskii}
\affiliation{\affiSOLAB}

\title{Polarimetric observation of   noise of linear birefringence   of cesium atomic vapors spectrally localized at double Larmor frequency   in a magnetic field.
 }
\begin{abstract}
We study radio-frequency spectrum of ellipticity noise  of a probe laser beam transmitted through a cell with cesium vapor in a magnetic field. The experimental results are interpreted in terms of the model proposed by Gorbovitskii and Perel (Opt. Spektrosc. {\ bf 54}, 388 (1983)) according  to which the observed noise arises due to heterodyning of the light scattered by fluctuations of the tensor $ \alpha $ of optical susceptibility of cesium vapor.  We show, both experimentally and theoretically, that, in the noise measurements if this kind, along with fluctuations of the antisymmetric (gyrotropic) part of the tensor $ \alpha $ at the Larmor frequency, may be observed fluctuations of its symmetric part, corresponding to fluctuations of linear birefringence (alignment) of the atomic system. The polarization noise provided by these fluctuations is localized spectrally at the double Larmor frequency.    
\end{abstract}

\maketitle

\section*{Introduction}

Studying the effects of scattering (interactions of particles or waves with a material) is one of the most important tools of physical experiment.  Reserford’s experiments on scattering of $\alpha$-particles that have shown planetary structure of atoms  analysis of the X-ray scattering that provided key information about structure of amorphous, crystalline, and quasi-crystalline materials may serve, among many others, as examples of application of the method of scattering in physics.  
The spectroscopy of spin noise (referred to, nowadays, as spin noise spectroscopy, SNS)  emerged and rapidly developing during the last decades, implies, in fact,  observation of polarization fluctuations of the probe laser beam scattered by a nonstationary, spatially inhomogeneous medium, with its circular birefringence (gyrotropy) oscillating in the applied magnetic field at the Larmor frequency $\omega_L\equiv g\mu B/\hbar$ (here $g$ is g-factor of the particles, contributing to the optical susceptibility detected in the SNS,  $\mu$ is the Bohr magneton, and $B$ is the applied magnetic field).

Recall basic principles of the SNS and briefly consider the main results  obtained with the use of this experimental technique. 
In the SNS, we detect radio-frequency spectrum of polarization noise of the light beam transmitted through (or reflected from) the sample under study.  The noise thus detected is related to fluctuations of the optical susceptibility tensor $\alpha$  of the sample. Most frequently, the SNS measurements imply detection of fluctuations of the anti-symmetric (gyrotropic ) part of the tensor $\alpha$, determined by {\it spin} states of the particles, which justifies the name of this particular kind  of the  light intensity fluctuations  spectroscopy \cite{Noise}. 

Magnetization of a medium is known to be connected with its spin state, with the noise spectrum of the magnetization, in accordance with the fluctuation-dissipative theorem, being determined by frequency dependence of the magnetic susceptibility of the sample. This is why, the spectra detected in the typical SNS experiments are, in essence, the spectra of imaginary part of magnetic susceptibility of the system,  which allows one to consider SNS as a version of the EPR spectroscopy.  Note, in this connection, that the first experiment on SNS  \cite{Zap}, performed in 1981, represented observation of the EPR spectrum of sodium atoms in the polarization noise (Faraday rotation noise) of a laser beam transmitted through the cell with sodium vapor. 

An important feature of the SNS is that it implies detection of the signals spontaneously generated by the sample, and when the wavelength of the probe beam corresponds to the transparency region of the sample, this kind of spectroscopy can be considered as {\it nonperturbative} . 

For the last  years, the SNS have shown itself as an efficient method of research  with a number of unique features (see reviews ~\cite{Zap1,Oest2,Sin}).  In particular, the SNS was used to detect and study the resonant magnetic susceptibility of quantum wells  and quantum dots in microcavities  that cannot be measured by the methods of conventional EPR spectroscopy ~\cite{Glazov,singlehole}. In ~\cite{R2}, a nonlinear instability of a semiconductor microcavity was studied and manifestations of the nuclear spin dynamics in SNS  in the above nanostructures were studied \cite{R,R1}.

It was found that by measuring dependence of the polarization noise power on the probe beam wavelength under conditions of optical resonance, it is possible to distinguish homogeneously and inhomogeneously broadened lines of optical transitions~\cite{Zap2}. The use of ultrashort laser pulses as a probe made it possible to expand the frequency range of the SNS to the region of microwave frequencies~\cite{To1}.
The two-beam version of the SNS proposed in~\cite{Koz} allows one to observe not only temporal, but also spatial correlations of the magnetization.  In~\cite{To}, there has been proposed a SNS-based method of magnetic tomography. This list of fields of application of the SNS and of the objects of this technique is not full and will be, undoubtedly, extended.    
 
In spite of the fact that in most SNS experiments the observed noise signals may be interpreted as magnetization noise of the sample under study, these signals are still detected in the {\it optical} channel with the use of {\it optical} photodetectors and therefore, strictly speaking, should be treated as a result of scattering of the probe laser beam by the sample.  This is why, a consistent treatment of the noise signals detected in the SNS should represent calculation of the probe beam by the medium with a fluctuating optical susceptibility.  The fact that the first experiments on SNS  \cite{Zap} may be interpreted as Raman scattering of the probe beam was pointed out in publication~\cite{Gorb}, whose ideas developed are here. 

The paper is organized as follows.  In Sec, “Experimental”, we present schematic of the experimental setup, describe the results of studying the noise spectra of cesium atoms under different experimental conditions  (for detecting the noise of Faraday rotation and ellipticity,  for different azimuths of the polarization plane, and for different intensities of the probe light).  This data is borrowed from  the  paper \cite {pr} of the authors.  In Sec. “Theoretical treatment”,  the theory of formation of the detected polarization noise is developed and it is shown that the signal at the double Larmor frequency results from the fact that the polarization noise detected in the SNS reveals not only fluctuations of gyrotropy of the atomic system (fluctuations of {\it orientation}), but also fluctuations of its linear anisotropy (fluctuations of {\it alignment}).  It is the last mentioned fluctuations that cause appearance of the peak at the double Larmor frequency $2\omega_L$ in the polarization noise power spectrum. In this section we also calculate orientational dependences of the noise signals at the frequencies $\omega_L$ and  $2\omega_L$.  In Sec. “Discussion” we show relationship between our theoretical results and experimental data and present general formula  for correlation function of the polarimetric noise that takes into account the Doppler broadening  and the time-of-flight effects. In Sec. “Conclusion”, we briefly summarize the results of the work. 

\section{\label{sec:Two}Experimental}

\begin{figure}
\includegraphics[width=9cm]{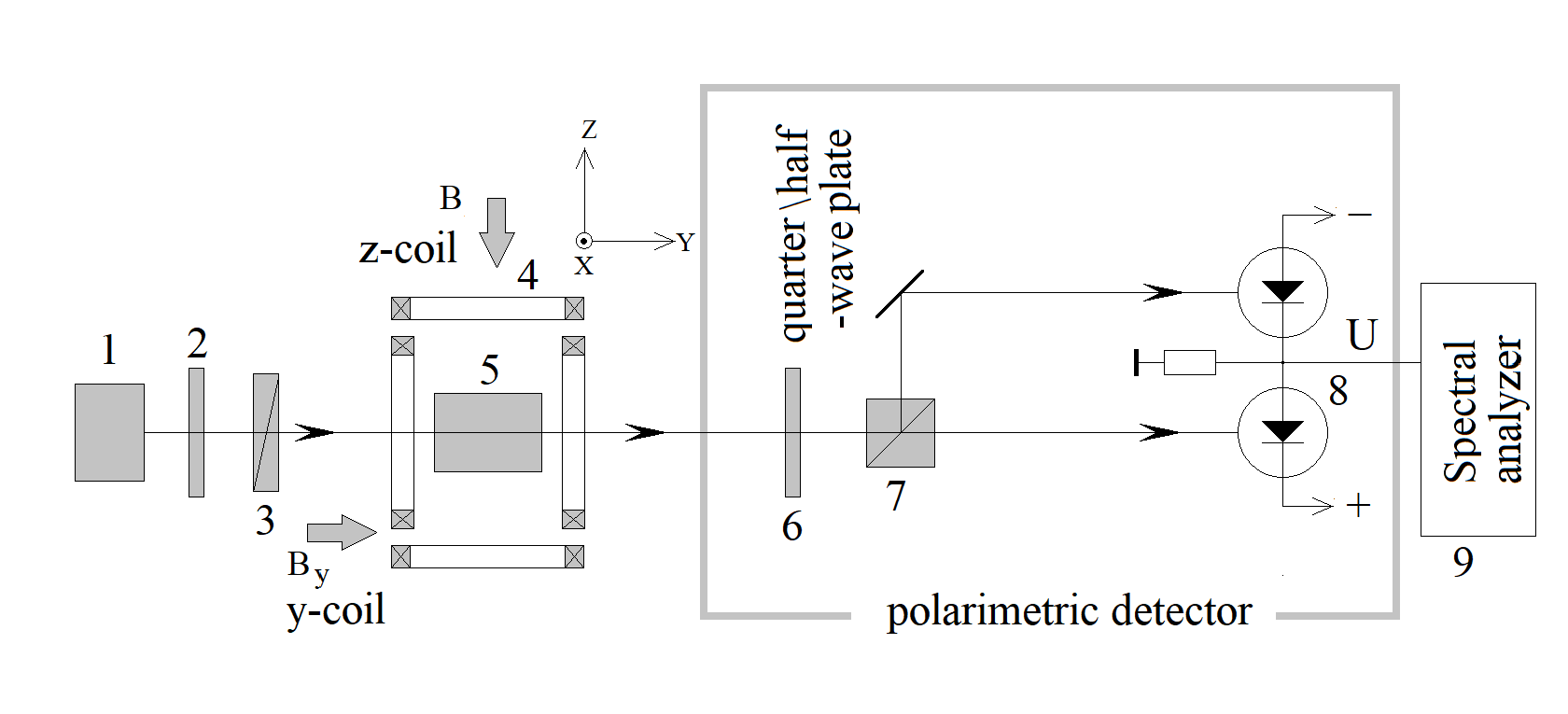}\\
\caption{Schematic of the experimental setup. 1 - laser, 2 - quater-wave plate, 3 - linear polarizer, 4 - magnetic coil, 5 - cesium vapor cell, 6 - phase plate, 7 - polarization beamsplitter, 8 -  differential photodetector, 9 - digital spectrum analyzer.}
\label{fig1}
\end{figure}

Schematic of the experimental setup is shown in Fig.~\ref{fig1}. Laser (1), quarter-wave plate (2), and linear polarizer (3) are used to prepare linearly polarized probe beam with its azimuth controlled by polarizer (3). The probe beam, after passing thorough cell  (5) with cesium vapor hits the polarimetric detector comprised of polarization beamsplitter (7) and differential photodetector (8).  In our experiments, the polarimetric detector could work in two regimes: in the regime of detection of the Faraday rotation noise and in the regime of detection of the ellipticity noise.  When detecting fluctuations of the Faraday rotation, the phase plate (6) was taken half-wave and was used for balancing the differential photodetector. When detecting fluctuations of ellipticity,  the phase plate (6) was taken quarter-wave, with its axes aligned at 45$^o$ with respect to axes of the beamsplitter (7).  For this arrangement of the polarization elements (as can be shown by direct calculations), the output signal of the differential photodetector is equal to zero, if the input light beam is linearly polarized (regardless of the polarization plane azimuth) and becomes nonzero only upon appearance of ellipticity in the input beam. 
The output electric signal of the differential photodetector (8) was fed to a digital spectrum analyzer (9) whose monitor displayed the polarization noise spectrum of the probe beam transmitted through the cesium cell (5). Most of our experiments were performed in the Voigt geometry with the light beam propagating across the magnetic field $B$ created by coil (4). Figure ~\ref{fig1} also shows the coil that created the magnetic field $B_y$ directed along the probe light propagation and used to perform measurements in the Faraday geometry.

\begin{figure}
	\includegraphics[width=9cm]{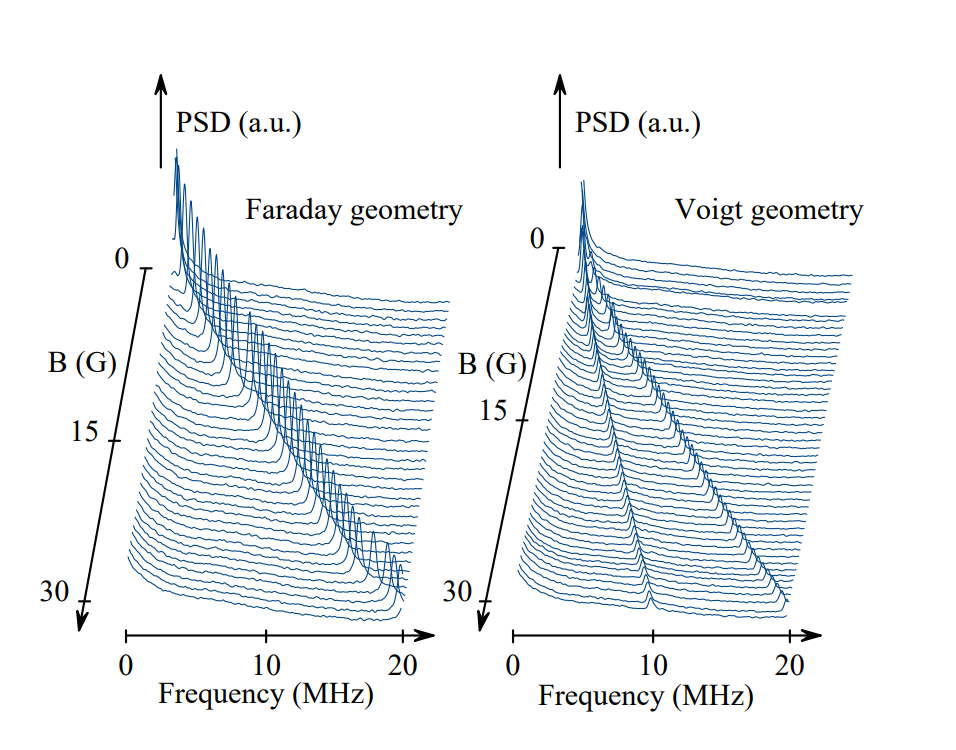}

	\caption{  Magnetic-field dependences of the ellipticity-noise power spectrum of Cs in the Voigt (right) and Faraday (left) geometries. In the Voigt geometry, one can see spin-noise peaks at the Larmor and double-Larmor frequencies, while in the Faraday geometry, only peak at the double Larmor frequency (spin alignment noise) is observed.  }
	\label{fig2}
\end{figure}

\begin{figure}[t]
	\subfigure{\label{fig2_a}}
	\subfigure{\label{fig2_b}}
	\subfigure{\label{fig2_c}}
	\includegraphics[width=\columnwidth,clip]{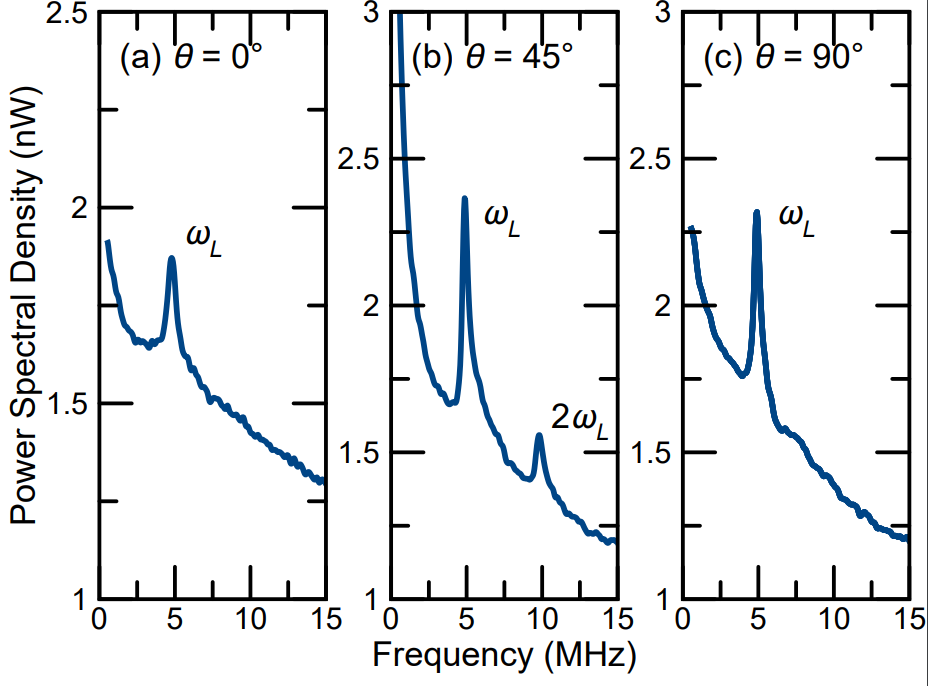}
	\caption{Ellipticity noise spectra detected at different angles $\theta$ between the probe beam polarization plane and magnetic field. The second harmonic of the Larmor frequency (at $\sim$ 10~MHz) is seen to be well pronounced at $\theta=45^o$ [panel (b)] and is not observed at $\theta= $0 and 90$^o$ [panels (a) and (c)]. The probe beam power density is $\sim 30$~mW/cm$^2$.}
	\label{fig3}
\end{figure}

 As it was already mentioned, the experiments were carried out with a cell with cesium vapor. The frequency of a probe beam  was close to the $ D2 $ cesium absorption line  ($ \lambda = 852.3 $ nm), and the ellipticity noise was observed in the Voigt and Faraday geometry. 
These experiments showed that the ellipticity noise spectra recorded in the Voigt geometry exhibited, along with the usual peak at the Larmor frequency $ \omega_L $ (observed in typical SNS experiments), also  a peak at double Larmor frequency $ 2 \omega_L $ (Fig.~\ref {fig2}) \cite {ff}.
The double Larmor frequency peak amplitude 
  reaches a maximum at an angle $ \theta $ between the directions of the magnetic field and the linear polarization of the probe beam equal to $ 45^o $ (Fig. ~ \ref {fig2_b}) and
vanishes at $ \theta = 0 $ and $ \theta = 90^o $ (see Figs. ~ \ref {fig2_a} and \ref {fig2_c}).
The ellipticity noise observed in the Faraday geometry reveals only one peak at the double Larmor frequency $ (2 \omega_L) $ (Fig.~\ref {fig2}, left).
 A study of the dependence of the described noise signals on the probe beam intensity showed that the double Larmor frequency peak is not a consequence of any nonlinear optical effect \cite {pr}. In the following sections, 
   we build up a theory  that describes  appearance of the indicated features of polarization noise at the frequencies $ \omega_L $ and $ 2 \omega_L $.  Our  calculations are  based on the fact that the observed signals (we call them the {\it noise signals}) are the result of scattering of the probe beam by fluctuations of the   linear optical  susceptibility tensor $ \alpha $ of the system of cesium atoms.

\section{\label{sec:Three}Theoretical treatment}

The polarimetric noise signal detected in our experiments is considered to be a result of scattering of the probe beam on cesium atoms~\cite{Gorb}. 
We will calculate this signal using the following simplifying assumptions:
(i) The electromagnetic field acting upon each atom, with an acceptable accuracy, coincides with that of the probe beam (approximation of single scattering);
(ii) Atomic polarization can be calculated in the approximation of linear response; 
(iii) The magnetic field is so small that Zeeman splitting of the atomic multiplets $\sim\omega_L$ is much smaller than the homogeneous linewidth $\delta$ and is not optically resolved ($\omega_L \ll \delta$). 

It is noteworthy that, strictly speaking, assumptions i) and ii) are not well satisfied: the resonant laser beam probing the cesium vapor exhibits substantial nonlinear absorption. 
Still, in the framework of the treatment presented below it appears possible to explain appearance of the second harmonic of the Larmor frequency in the polarization noise spectrum and to qualitatively describe its main properties.  

\subsection{Calculation of the polarimetric signal}

In our experiments, the noise signal at the double Larmor frequency was most pronounced in the \emph{ellipticity} noise spectra. 
This is why, in what follows, we present calculations for the noise signal of this kind (the Faraday rotation noise spectrum can be calculated in a similar way~\cite{Koz}). 
In this case, the quarter-wave plate of the polarimetric detector [Fig.
~\ref{fig1}]
 is aligned with its axes at $45^\circ$ with respect to polarization directions of the PBS, which we assign to be the axes $z$ and $x$.  
The probe beam propagation direction is taken for the axis $y$ [see Fig.
~\ref{fig1}]
In this coordinate system, the magnetic field has only $z$~component, ${\bf B}=(0,0,B)$, while the probe-beam electric field $\bf {\cal E}_0$ has only the $x$ and $z$~components, ${\bf {\cal E}}_0=({\cal E}_{x0},0,{\cal E}_{z0})$ [Fig.
~\ref{fig1}]

Let us denote electric field  of the probe beam at the input of the polarimetric detector as ${\bf {\cal E}}=({\cal E}_{x},{\cal E}_{y},{\cal E}_{z})$. 
Then, as can be shown by direct calculations, the output signal $U$ of the detector, in this mode of operation, is given by 
\begin{equation}
U={1\over T}\int_0^Tdt \int_S dx dz \bigg [ {\cal E}_x(t){\cal E}_z(t+\Delta) - {\cal E}_x(t+\Delta) {\cal E}_z(t)\bigg ] 
\label{1}
\end{equation} 
where $\Delta={\pi\over 2\omega}$ and $\omega$ is the probe light frequency.
 The integration over $dxdz$ in Eq.~\eqref{1} is performed over photosensitive surfaces of the photodetectors $S$, which are supposed to be identical. 
The integration over $dt$ corresponds to averaging over the time interval $T$ that contains integer number of optical periods and meets the requirement $2\pi/\omega \ll T \ll 2\pi/ \omega_L$. 
We see that, indeed, the signal $U$ appears to be nonzero only for the elliptically polarized input field ${\bf {\cal E}}_0$, while for any linearly polarized field, the output signal vanishes.  
It  is also seen from Eq.~\eqref{1} that any rotation of the polarimetric detector around the $y$~axis does not affect the output signal $U$, since for any two vectors $\bf A$ and $\bf B$, the quantity 
\begin{equation}
A_xB_z-A_zB_x= ({\bf A,\beta B}),\hskip3mm\beta\equiv \begin{pmatrix} 0 & 1 \\ -1 & 0 \end{pmatrix} 
\label{metric}
\end{equation}
does not change under arbitrary rotations in the plane $xz$.    

The noise signal observed in our experiments can be represented as the sum of the contributions of individual atoms. Therefore,
let us now calculate the ellipticity signal $\delta u_e$, created by a single atom. 
For convenience, we will consider the field $\cal E$ at the input of our detector as a real part of the complex field ${\bf E}$: ${\cal E}= \mathop{\rm Re} {\bf E}$. 
The field ${\bf E}$ can be considered as a sum of the field ${\bf E}_0\equiv {\bf A}_0 e^{-\imath\omega t}$ of the probe beam $({\cal E}_0= \mathop{\rm Re} {\bf E}_0)$ and the field ${\bf E}_1$ created by the atomic dipole $({\cal E}_1= \mathop{\rm Re} {\bf E}_1)$. 
We use calligraphic letters to denote the observed (real) fields.

It suffices to calculate the ellipticity signal in the approximation linear in the field ${\bf E}_1$. 
Since all the fields are assumed  quasi-monochromatic ($\sim e^{-\imath\omega t}$), the time shift by $\pm \pi/2\omega$ is equivalent to multiplication by $\mp \imath$. 
Keeping this in mind, we obtain from Eq.~\eqref{1} the following expression for the signal $\delta u_e$ \cite{for}
\begin{equation}
\delta u_e=   \hbox{ Im }
{2\over T}\int_0^Tdt \int_S dx dz \bigg [ {\cal E}_{0x}{ E}_{1z}  -{E}_{1x} {\cal E}_{0z} \bigg ].
\label{2}
\end{equation}
This formula includes both complex $(E_{1i})$ and real fields $({\cal E}_{0i})$.
The field ${\bf E}_1$ of the atomic dipole can be obtained by solving the inhomogeneous Helmholtz equation $\Delta {\bf E}_1 +k^2{\bf E}_1=-4\pi k^2{\bf P}$, where $k=\omega /c$ ($c$ is the speed of light) and $\bf P(r)\sim \delta ({\bf R-r})$ is the complex polarization created by the atom with the radius-vector $\bf R$.  
Solution of this equation can be obtained using Green's function of the Helmholtz operator (see, i.g.,~\cite{Koz,Koz1}) and has the form ${\bf E}_1({\bf r})=k^2\int d^3{\bf r'}e^{\imath k|{\bf r-r'}|}{\bf P(r')}/|{\bf r-r'}|$. 
By substituting this expression into Eq.~\eqref{2}, we have:

\begin{equation}
\delta u_e={2k^2\over T} \hbox{ Im } \int_0^Tdt \int d^3{\bf r'} \bigg [
\Phi_x({\bf r'})P_z({\bf r'})-\Phi_z({\bf r'})P_x({\bf r'})
\bigg ],
\label{3}
\end{equation}
where we introduced the following functions $\Phi_i({\bf r'}) (i=x,z)$\cite{f0}:
\begin{equation}
\begin{split}
\Phi_i({\bf r'})\equiv \int _S dxdz \hskip1mm {\cal E}_{0i}(x,y,z){e^{\imath k |{\bf r-r'}|}\over |{\bf r-r'}|}\bigg |_{r_y=L} \equiv\\ 
e^{-\imath\omega t}\Phi_i^+({\bf r'})+e^{\imath\omega t}\Phi_i^-({\bf r'}), \hskip3mm i=x,z,
\hskip3mm {\bf r}=(x,y,z)
\label{4}
\end{split}
\end{equation}

When integrating over the photodetector surface $S$ ($dxdz$) in Eq.~\eqref{4}, we assumed that $r_y=y=L$, where $L$ is the distance from the atom to the polarimetric detector, which we consider to be large: $L\rightarrow \infty$. 
Besides, in Eq.~\eqref{4} we separate explicitly the components $\Phi_i^\pm({\bf r'})$ of the function $\Phi_i({\bf r'})$ proportional to $e^{\mp\imath\omega t}$. 
In the approximation of linear response, atomic polarization is proportional to the probe wave electric field. 
Therefore, ${\bf P(r')}\sim e^{-\imath\omega t}$ and, after time-averaging in Eq.~\eqref{3}, only components $\Phi_i^-({\bf r'})$ survive. 
In~\cite{Koz1}, it has been shown that 
\begin{equation}
\Phi_i^-({\bf r'}) =-{\imath \pi\over k} A_{0i}^\ast({\bf r'}), \quad
|{\bf r'}|\ll L, \quad i=x,z
\label{5}
\end{equation}  
where $A_{0i}$ is the $i$-th projection of the amplitude of the probe beam complex field.
The polarization ${\bf P(r')}$ created by a single atom entering Eq.~\eqref{3} can be presented in the form  $P_i({\bf r'})=\delta({\bf r'-R})\langle d_i\rangle e^{-\imath\omega t}$ where $\langle d_i\rangle $ is the complex amplitude of oscillation of the $i$-th component of the atomic dipole moment, $\bf R$ is the radius-vector of the atom.  
By substituting this expression into Eq.~\eqref{3} and taking into account Eq.~\eqref{5}, we obtain, for the ellipticity signal $\delta u_e$ created by a single atom, the following expression:
\begin{equation}
\delta u_e = 2\pi k\hbox{ Re }\bigg [
A_{0x}^\ast({\bf R})\langle d_z\rangle -A_{0z}^\ast({\bf R})\langle d_x\rangle
\bigg ].
\label{6}
\end{equation} 

When the polarimetric detector operates in the Faraday-rotation detection mode (i.e., the $\lambda/4$ wave plate is replaced by the $\lambda/2-$ plate), then a similar calculation leads to the following expression for the Faraday rotation noise signal $\delta u_r$ produced by a single atom:
\begin{equation}
\begin{split}
\delta u_r= 2\pi k \mathop{\rm Im} \bigg [\cos [2\phi]\bigg (A_{0x}^\ast({\bf R})\langle d_x\rangle- 
A_{0z}^\ast({\bf R})\langle d_z\rangle\bigg )  -\\
\sin [2\phi]\bigg (A_{0x}^\ast({\bf R})\langle d_z\rangle + A_{0z}^\ast({\bf R})\langle d_x\rangle\bigg )\bigg ]
\label{7}
\end{split}
\end{equation} 
Here $\phi$  is the angle between $z$-axis and one of the main directions of the beamsplitter (In the above calculation of the ellipticity signal, we used a coordinate system for which $ \phi = 0 $. See \cite{fbs} for explanation).
 
 Equations \eqref{6} and \eqref{7} can be simplified under the following conditions. 
First, we assume that the quantities $\langle d_i\rangle$ entering Eqs.~\eqref{6} and \eqref{7} can be expressed through the probe light electric field ${\bf A}_0({\bf R})$ using the susceptibility tensor $\alpha$: $\langle d_i\rangle =\alpha_{ik}A_{0k}$. 
Second, the probe beam is assumed to be linearly polarized. 
In this case, $A_{0x}=A_0\sin\theta$ and $A_{0z}=A_0\cos\theta$, where $\theta$ is the angle between the probe beam polarization and $z$ axis. 
And, third, we assume that, in the measurements of the Faraday-rotation noise, orientation of the polarimetric detector (specified by the angle $\phi$ or, which is the same, by the orientation of the half-wave plate, see \cite{fbs}) corresponds to conditions of balance with no DC signal at the output of the detector, i.e., $\phi=\theta+\pi/4$ \cite{f11}.
When the above conditions are satisfied, Eqs.~\eqref{6} and \eqref{7} can be rewritten in a  compact  scalar form as follows 
\begin{equation}
\begin{split}
\delta u \equiv \delta u_e+\imath \delta u_r=
2\pi k({\bf A}_0({\bf R}),\beta\alpha {\bf A}_0({\bf R}))= \\ 
\pi k|A_0({\bf R})|^2\bigg [
\alpha_{zx} - \alpha_{xz} - 
(\alpha_{xz}+\alpha_{zx})\cos 2\theta + \\
(\alpha_{zz}-\alpha_{xx})\sin 2\theta
\bigg ]
\label{8}
\end{split}
\end{equation}
 with matrix $\beta$  defined by Eq.\ref{metric}.

\subsection{Calculation of the atomic susceptibility}

Calculation of the linear atomic susceptibility is performed assuming that it is related to optical transitions between two (ground and excited) atomic multiplets \cite{f12}    
with the same total angular momenta $F$.  
Formation of the optical response of the atom flying into the probe beam can be imagined in the following way. 
The wavefunction of the atom $\Psi(0)$, at the moment of its entering the beam (let it be $t=0$) is a random superposition of atomic eigenfunctions $|1M\rangle$ of the ground multiplet with different $z$~components ($M$) of the angular momentum: 
$\Psi(0)=\sum_{M=-F}^FC_M|1M\rangle=\sum_{M=-F}^F|C_M|e^{\imath \beta_M}|1M\rangle$, 
where $C_M$ and $\beta_M$ are the random amplitude of the atomic state $|1M\rangle$ and its phase (with $\sum_{M=-F}^F|C_M|^2=1$). 
When the magnetic field is nonzero, the ground multiplet exhibits Zeeman splitting and the above superposition state appears to be nonstationary (even neglecting the probe beam induced perturbation). 
The appropriate unperturbed density matrix of the atom $\rho_0$ also appears to be time-dependent and has nonzero matrix elements only in the subspace of the states of the ground atomic multiplet:   
\begin{equation}
\langle 1M| \rho_0(t)|1M'\rangle=
|C_M||C_{M'}|e^{\imath [\beta_M-\beta_{M'}]} e^{\imath \omega_{1L}[M-M']t}
\label{9}
\end{equation}
where $\omega_{1L}$ is the Larmor frequency for the ground-state multiplet. 
As seen from Eq.~\eqref{9}, the density matrix oscillates in time at frequencies integer multiples of the Larmor frequency $\omega_{1L}$. 
Since the linear optical susceptibility of the atom is related to its unperturbed density matrix (this connection will be presented below), it may depend on time at frequencies  $\omega_{1L}[M-M']$. 
This may, in turn, give rise to appearance of shifted frequencies $\omega\pm \omega_{1L}|M-M'|$  in the spectrum of the field ${\bf E_1}$ scattered by the atom (the effect of Raman scattering) and can be detected in our experiments as the noise of polarimetric signal spectrally localized in the vicinity of the frequencies  $\omega_{1L}|M-M'|$.  
 As will be seen below, only frequencies with $|M-M'|=0,1,2$  can be observed
and, correspondingly, only spectral features at the frequencies $0$, $\omega_{1L}$, and $2\omega_{1L}$ can arise in the polarization noise spectra.  

Let us pass now to calculation of the linear atomic susceptibility. 
We perform calculations for the case of Voigt geometry, the case of Faraday geometry can be analysed in a similar way.
The matrix of the Hamiltonian of the atom in the representation of the two (ground and excited) multiplets in frequency units has the form
\begin{equation}
\begin{split}
H=H_0+H_E, \\
H_0 \equiv \Omega \begin{pmatrix} I & 0\\ 0 &0 \end{pmatrix} + \begin{pmatrix} \omega_{2L} J_z & 0\\ 0 & \omega_{1L}J_z \end{pmatrix},\\
H_E= \omega_{x}e^{-\imath\omega t} \begin{pmatrix} 0 & J_x \\ J_x & 0 \end{pmatrix}
+\omega_{z}e^{-\imath\omega t} \begin{pmatrix} 0 & J_z \\ J_z & 0 \end{pmatrix}
\label{10}
\end{split}
\end{equation}
where $\omega_{iL}$ is the Larmor frequency of the $i$-th multiplet ($i=1,2$) and the Rabi frequencies $\omega_{x,z}$ are determined by the dipole moment ($d$) of the atomic transition between the multiplets  and by projections of the amplitude of the probe field at point  $\bf R$ where the atom is located: $\omega_{i}\equiv dA_{0i}({\bf R})/ \hbar, i=x,z$. 
Each `element' of matrices in Eq.~\eqref{10} is itself a matrix with dimensions $(2F+1)\times(2F+1)$, with $J_z$ and $J_x$ being known matrices of the corresponding projections of the angular momentum $F$ \cite{Landau}. 
The matrices of the operators for the needed $x$ and $z$~projections of the atomic dipole moment have the form

\begin{equation}
d_x\equiv d \begin{pmatrix} 0 & J_x \\ J_x & 0 \end{pmatrix}, \quad
d_z\equiv d \begin{pmatrix} 0 & J_z \\ J_z & 0 \end{pmatrix}.
\label{11}
\end{equation}

The standard procedure of the linear-response theory  implies representation of solution of the equation $\imath \dot \rho=[H,\rho]$ for the atomic density matrix $\rho$ in the form $\rho=\rho_0+\rho_1+O(H_E^2)$ where $\imath\dot \rho_0=[H_0,\rho_0]$ and $\imath\dot\rho_1=[H_0,\rho_1]+[H_E,\rho_0]$ and computation of the quantities $\langle d_i\rangle$ as  $\langle d_i\rangle = \mathop{\rm Sp}\rho_1d_i$ where $i=x,z$. It leads to the expression $\langle d_i\rangle =\alpha_{ik}A_{0k}({\bf R})$, in which the susceptibility tensor $\alpha$ contains the following elements
\begin{equation}
\begin{split}
\alpha_{ik}\equiv {d^2\over \hbar }\sum_{MM'M''}{\langle 1M|\rho_0|1M'\rangle \langle M'|J_k|M''\rangle \langle M''|J_i|M\rangle\over \Delta\omega+\imath\delta + \omega_{2L}M''-\omega_{1L}M'  },\\ 
i,k=x,z
\end{split}
\label{12}
\end{equation}

Here, $\Delta\omega\equiv \Omega-\omega$ is the optical detuning, $\imath\delta$ denotes the homogeneous broadening, $\langle M|J_k|M'\rangle$ are the matrix elements of the operator of $k$-th projection of the angular momentum $F$~\cite{Landau}, and the summation over $M,M'$, and $ M''$ is performed over $2F+1$ states of the ground-state multiplet. 
As has been noted above, the tensor $\alpha$ depends on time (through the matrix elements $\langle 1M|\rho_0|1M'\rangle $, see Eq.~\eqref{9}), with characteristic frequencies of this dependence corresponding to spectral features of the noise spectra observed in the SNS. 

Since the quantities $\langle M|J_z|M'\rangle =\delta_{MM'}M$ and $\langle M|J_x|M'\rangle$ are nonzero only for $|M-M'|= 1$, it follows from Eq.~\eqref{12} that the only frequencies $\omega_{1L}|M-M'|$ at which oscillations of the tensor $\alpha$ may occur are:
$\omega_{1L}$ (the elements $\alpha_{xz},\alpha_{zx}$ $\sim \langle M'|J_x|M''\rangle\langle M''|J_z|M\rangle$),   
$2\omega_{1L}$ (the elements $\alpha_{xx}$ $\sim \langle M'|J_x|M''\rangle\langle M''|J_x|M\rangle$) and 0 (the elements $\alpha_{zz},\alpha_{xx}$ $\sim \langle M'|J_x|M''\rangle\langle M''|J_x|M\rangle, \langle M|J_z|M''\rangle\langle M''|J_z|M\rangle) $). 

It is seen from Eq.~\eqref{8} for the complex polarimatric signal $\delta u$ that the components of this signal at the frequency $2\omega_L$ behave as $\sim \sin 2\theta$ and vanish when the probe beam polarization is parallel or perpendicular to the magnetic field ($\theta=0,\pi/2$), as it is observed in our experiments (Fig.~\ref{fig2}). 

\subsection{Calculation of the polarimetric signal} 

Note that under assumption that $\omega_{1L}\ll\delta$ the dependence of the denominator in Eq.~\eqref{12} on the numbers $M''$ and $M'$ may be neglected. 
Then we obtain the following expression for the matrix of the atomic susceptibility $\alpha$ 
\begin{equation}
\alpha_{ik} ={d^2\hbox { Sp }\rho_0 J_kJ_i \over \hbar [\Delta\omega +\imath \delta]}
={d^2\hbox { Sp }\rho_0 [\{J_kJ_i\} +\imath\varepsilon_{kil}J_l]\over 2\hbar [\Delta\omega +\imath \delta]}, 
\label{13}
\end{equation}  
in which we selected symmetric ($\sim \{J_kJ_i\}\equiv J_kJ_i+J_iJ_k$) and antisymmetric (gyrotropic, $\sim J_kJ_i-J_iJ_k=\imath \varepsilon_{kil}J_l$) parts (here $\varepsilon_{kil}$ is Levi-Civita tensor).
After such a simplification, the expression for the susceptibility $\alpha$ acquires the form of a quantum mean value of a tensor observable with the operator $\sim J_kJ_i$ in the state with the density matrix $\rho_0$. 
The appropriate superpositional wavefunction $\Psi$ (it contains only the components related to the ground multiplet) satisfies the Schr\"odinger equation $\imath \dot \Psi=H_0\Psi = \omega_{1L}J_z\Psi$ and is defined by the formula: $\Psi(t)=e^{-\imath \omega_{1L}J_zt}\Psi(0)$. 
Since the operator $e^{-\imath \omega_{1L}J_zt}$ is the operator of rotation by the angle  $\omega_{1L}t$ around the $z$~axis~\cite{Landau}, the function  $\Psi(t)$ represents the function $\Psi(0)$, rotating around the magnetic field with the angular frequency $\omega_{1L}$. 
This rotation is accompanied by `rotation' of the tensor  $\alpha_{ik}\sim \mathop{\rm Sp} \rho_0 J_kJ_i= \langle \Psi |J_kJ_i|\Psi\rangle$ \cite{f13}, and the noise signal detected in our experiments can be understood as a result of scattering of the probe beam by a quasi-point anisotropic system rotating with the Larmor frequency $\omega_{1L}$ around the magnetic field.  
If we substitute Eq.~\eqref{13} into  \eqref{8}, \textit{}we obtain for the complex polarimetric signal $\delta u$ the following expression
\begin{equation}
\delta u= {\pi kd^2\over \hbar }|A_0({\bf R})|^2 f(t),
\label{14}
\end{equation}
where 
\begin{widetext}
	\begin{equation*}
	f(t) \equiv {\mathop{\rm Sp}\rho_0\bigg [ (J_x^2-J_z^2)\sin 2\theta +(J_zJ_x+J_xJ_z)\cos 2\theta +\imath J_y 
		\bigg ]\over \Delta\omega+\imath \delta}\equiv f_e(t)+\imath f_r(t)
	\end{equation*}
\end{widetext}

Physical meaning of different contributions in this formula can be determined by considering behavior of the function $f(t)$ at large detunings $\Delta\omega\gg\delta$. 
It can be seen that the first two terms in Eq.~\eqref{14} describe fluctuations of symmetric part of the tensor $\alpha$ (fluctuations of \emph{alignment}) and, being real (at $\Delta\omega\gg\delta$), can be observed only in the regime of detection of ellipticity (see Eq. (\ref{8})). 

Since the matrix elements $\langle M|J_x^2-J_z^2|M'\rangle$ are nonzero only at $|M-M'|=0$ and $|M-M'|=2$, the contribution $\sim (J_x^2-J_z^2)\sin 2\theta$ gives rise to peaks in the spectra of ellipticity noise at zeroth and double Larmor frequencies.
 (Since they enter the expression for the polarimetric signal Eq. (\ref{14}) together with the elements $\langle M'|\rho_0|M\rangle$ of the density matrix whose time behaviour is detrmined by Eq. (\ref{9}).
 
In a similar way, one can make sure that the contribution $\sim (J_zJ_x+J_xJ_z)\cos 2\theta$ gives rise to a feature at the frequency $\omega_{1L}$. 
The constant (isotropic) term in brackets $\sim \imath \mathop{\rm Sp} \rho_0 J_y$ describes fluctuations of gyrotropy of the atomic system and, being pure imaginary, is revealed only in the Faraday rotation noise. 
Since the matrix elements $\langle M|J_y|M'\rangle$ are nonzero only at $|M-M'|=1$, this term provides a feature in the Faraday-rotation noise spectrum only at the frequency $\omega_{1L}$.

  The above consideration was related to the case of Voigt geometry. Similar results can be obtained for the Faraday configuration. In this case, the expression for the complex polarimetric signal $ \delta u $ differs from Eq. (\ref {14}) by the permutation of the operators
  $ J_z \rightarrow J_y $ and $ J_y \rightarrow J_z $ (leaving the same  expression Eq. (\ref {9})  for the density matrix $ \rho_0 $).   An analysis similar to the above shows that the  polarimetric noise signal   recorded in Faraday geometry will have spectral features only at zero frequency and at a frequency of $ 2 \omega_ {1L}$.
  
    The rigour calculation of the noise power spectrum ${\cal N}(\nu)=\int \langle f(0)f(t)\rangle e^{-\imath\nu t}dt$ observed in our experiments requires
    the calculation of  the correlation function $\langle f (0)f (t)\rangle $.
   This calculation is somewhat cumbersome. Below (in section \ref{sec:Dis} ) we present the results of such calculation with no details 
    which will be published elsewhere at the request of readers. 
     The quantum-mechanical correlation functions of the operators entering Eq. (\ref{14}) were calculated in \cite{pr}.
   
   
   
  
\section{\label{sec:Dis} Discussion}
The above simplified consideration shows that observation of spectral feature at the frequency $2\omega_{1L}$ is possible only in the ellipticity noise spectrum. 
Remind that our experiments mainly support this conclusion. 
A consistent calculation shows, however, that when the homogeneous width of the line $\delta$ is getting much smaller than the Doppler broadening, the difference between the noise spectra of ellipticity and Faraday rotation (in terms of the peak at the double Larmor frequency) becomes not so dramatic. 
It can be shortly explained as follows. 
Consider, e.g., the ellipticity noise spectrum, which is determined by the Fourier-image of the correlation function  $\langle \delta u_e(t)\delta u_e(0)\rangle\sim \langle |A_0({\bf R}(t))|^2|A_0({\bf R}(0))|^2 f_e(t)f_e(0)\rangle$ [see Eq.~\eqref{14}]. 
Calculation of correlator of the signal $f_e(t)$ \eqref{14} leads to the following expression:
\begin{widetext}
	\begin{equation}
	\begin{split}
	&\langle f_e(t)f_e(0)\rangle\sim \bigg \{ {5 a^2} \cos [\omega_{1L}t] +\\
	&{d^2}\bigg [F(F+1)-{3\over 4}\bigg ]\bigg [ 4\cos^22\theta \cos [\omega_{1L}t]+ (3+\cos [2\omega_{1L}t]) \sin^2 2\theta \bigg ]\bigg \}, \quad
	d+\imath a\equiv 1/[\Delta \omega+kv_y +\imath \delta]
	\end{split}
	\label{15}
	\end{equation} 
\end{widetext}      
A similar expression was obtained in the theoretical section of work \cite{pr} by solving the equations of motion for correlation functions.
Here we omitted not essential factors and accounted for the Doppler shift $kv_y$ ($v_y$ is the projection of the atomic speed upon the probe beam direction). 
The expression for the correlator of the Faraday rotation signal  $\langle f_r(t)f_r(0)\rangle $ differs from Eq.~\eqref{15} by the substitutions 
$a\rightarrow d$ and $ d\rightarrow a$. 
Despite the fact that frequency dependence of the correlators $\langle f_e(t)f_e(0)\rangle$  and $\langle f_r(t)f_r(0)\rangle $ is different, for both of them it has the form of a sharp feature with the width  $\sim\delta$. 
For this reason, upon Maxwellian averaging of the Doppler shift $kv_y$, with the width $\sim kv_T\gg \delta$ (here, $v_T^2$  is the mean-square thermal velocity), the above difference (for  $\Delta\omega \sim kv_T$) will be of no importance. 
For the Gaussian probe beam, calculation leads to the following expression for the correlation function observed in the SNS (inessential numerical factors are omitted):
\begin{widetext}
	\begin{equation}
	\begin{split}
	&K(t)\sim {\sigma  \hskip1mm  W^2 \rho_c k^2 d^4\hskip1mm
		\over  v_T^2 \delta }  
	\exp \bigg [-{\Delta\omega^2 \over k^2v_T^2} \bigg ]   \times \\
	&{e^{-|t|/T_2} \over \sqrt{t^2+t_T^2}} \bigg \{ 5 \cos [\omega_{1L}t] 
	+\bigg [F(F+1)-{3\over 4}\bigg ]\bigg [ 4\cos^22\theta \hskip1mm\cos [\omega_{1L}t]+ (3+\cos [2\omega_{1L}t])\hskip1mm
	\sin^2 2\theta \bigg ]\bigg \}
	\end{split}
	\label{16}
	\end{equation}
\end{widetext}

Here, along with the quantities introduced above, we use: $\sigma$ -- atomic vapor density, $W$ -- the probe beam power, $\rho_c$ -- the beam radius in its waist,  $t_T\equiv \rho_c/ v_T$ -- the time of flight, and $T_2$ -- spin relaxation time. 
The polarization noise power ${\cal N}(\nu)$ is defined as ${\cal N}(\nu)=\int K(t) e^{\imath \nu t} dt$. 
As seen from Eq.~\eqref{16}, at $T_2,t_T\gg  \omega_{1L}^{-1}$, the spectrum of the polarization noise power of atomic vapor always shows features at $\nu =0,\omega_{1L},$ and $ 2\omega_{1L}$. 
Angular dependence of amplitudes of these features at  $F>1/2$ is controlled by the last term in brackets and $\sim \cos^2 2\theta$ for the feature at $\nu=\omega_{1L}$ and $\sim \sin^2 2\theta$ for the feature at $\nu=2\omega_{1L}$. Recall once again that nonlinear effects were not taken into account in (\ref{16}).

Dependence Eq. (\ref{16}) of the ellipticity noise spectra on azimuth $\theta$ of the polarization plane of the probe beam qualitatively agrees with the experimental data [see Fig.~\ref{fig3}] -- amplitude of the peak at the frequency $2\omega_{1L}$ reaches maximum at $\theta=\pi/4$ and vanishes at $\theta=0$ and $\pi/2$.

\section{\label{sec:conclusion}Conclusion}

The suggested paper describes the mechanism of formation of polarimetric noise (ellipticity noise  and Faraday rotation noise), produced by atomic vapors in an external magnetic field and observed by means of spin noise spectroscopy (SNS) technique.

The observed noise signals are interpreted as a result of scattering of the probe beam by an atomic system, whose optical susceptibility  undergoes fluctuations. It is shown, that
   in  general case, the power spectrum of the polarimetric noise of an atomic system reveals features (maxima) at
    zero, first and second harmonics of Larmor frequency $\omega_L$.
According to our calculations, the unusual feature at  double Larmor frequency $2\omega_L$ is associated with fluctuations of the symmetric part of the tensor of atomic optical susceptibility (alignment fluctuations), in contrast to the feature at  Larmor frequency $\omega_L$ observed in typical SNS experiments and associated with fluctuations of the antisymmetric part of the optical susceptibility tensor (girotropy flactuations).
The calculated dependence of the noise spectrum on the angle between the directions of the magnetic field and polarization of the probe beam is in qualitative agreement with the  experiment \cite {pr}.

\begin{acknowledgements}
 This work was supported by the Russian Science Foundation (grant No. 17-12-01124). I.I.R. acknowledges President of the Russian
Federation Grant No. MK-2070.2018.2 for support of the
preliminary spectroscopic characterization of the cesium
vapor cells. The work was performed using equipment of
the SPbU Resource Center “Nanophotonics”.

The authors are grateful to M.M. Glazov for useful discussions.
	\end{acknowledgements}


\begin{references}
\bibitem{ff} Figures (\ref{fig2}), (\ref{fig3}) are borrowed from work \cite{pr}.
\bibitem{for} Upon receipt of this formula, generally speaking, four terms arise. However, proper replacement of the variable during integration over time and doubling of the result allows us to obtain Eq. (\ref{2}).
\bibitem{f0} The function $\Phi({\bf r})$ has the sense of the field created by the source distributed over the surface of the detector with the density ${\bf {\cal E}}_0(x,y,L)$. For this reason, the field $\Phi({\bf r})$ is similar to that of the main beam ${\bf {\cal E}}_0({\bf r})$. Appropriate formula is derived in \cite{Koz1} and is given below (Eq. (\ref{5})).
\bibitem{fbs} Rotatation  of the half-wave plate (6)  (see Fig.\ref{fig1}) by any angle $\xi$ is equivalent to changing $\phi$ by $\phi+2\xi$. Therefore, the polarimetric detector can be ballanced by rotating the half-wave plate (6) (see text below). For the  introduced above coordinate system with $z$-axis parallel to main direction of beamsplitter, we can set $\phi=2\xi$.
\bibitem{f11} The two last conditions were satisfied in our experiments.
\bibitem{f12} In reality, the hyperfine component of the cesium $D2$ line, we are interested in, is associated with transitions from the multiplet $F=4$ to the multiplets $F=3,4,5$, but for the semiquantitative interpretation of the observed effects this simplification seems acceptable.
\bibitem{f13} Mentioned rotation is the rotation of the principal axes of the symmetric part of the tensor $\alpha$ and the rotation of the gyration vector corresponding to its antisymmetric part.
\end{references}
 \end{document}